# Force-directed algorithms for schematic drawings and placement: a Survey

Se-Hang Cheong

Department of Computer and Information Science,
University of Macau
dit.dhc@lostcity-studio.com

Yain-Whar Si

Department of Computer and Information Science,
University of Macau
fstasp@umac.mo

Force-directed algorithms have been developed over the last 50 years and used in many application fields, including information visualisation, biological network visualisation, sensor networks, routing algorithms, scheduling, graph drawing, etc. Our survey provides a comprehensive summary of developments and a full roadmap for state-of-the-art force-directed algorithms in schematic drawings and placement. We classified the model of force-directed algorithms into classical and hybrid. The classical force-directed algorithms are further classified as follows: (a) accumulated force models, (b) energy function minimisation models, and (c) combinatorial optimisation models. The hybrid force-directed algorithms are classified as follows: (a) parallel and hardware accelerated models, (b) multilevel force-directed models, and (c) multidimensional scaling force-directed algorithms. Five categories of application domains in which force-directed algorithms have been adopted for schematic drawings and placement are also summarised: (a) aesthetic drawings for general networks, (b) component placement and scheduling in high-level synthesis of very-large scale integration (VLSI) circuits design, (c) information visualisation, (d) biological network visualisation, and (e) node placement and localisation for sensor networks.

## 1    INTRODUCTION

Force-directed algorithms have been developed over the last 50 years and adopted in numerous application fields. For example, these include: visualising genetic structures automatically in biology, optimising networks for parallel computer architectures, detecting clusters and hidden patterns in the social sciences, placing and scheduling components for very-large-scale integration circuits (VLSI), and computing undirected/directed networks (graphs) for information visualisation, etc. A schematic drawing is a representation of the elements of a network using simple graphic symbols. Such drawing shows crucial components of the network and the details that are not relevant to the information are omitted [1]. For example, a dot may be used to represent a station in a subway map. In this case, the dot is used to provide key location information to the users without causing any unnecessary visual cluttering. In some application domains, the size of the canvas and the detailed arrangement of the elements in the drawing are constrained by certain technical limits. Force-directed placement [2] is one of the approaches for the node placement in the schematic drawing. The placement of nodes along the edges or in a specific region of the canvas are useful in VLSI applications. According to statistics on annual paper submissions related to force-directed algorithms depicted in Figure 1 (a), force-directed algorithms are very popular and have often been preferred over other algorithms since the 1980s. Figure 1 (a) and Figure 1 (b) show a classification of force-directed algorithms by trends in paper submission and application fields. According to our review, 38% of force-directed algorithm studies relate to schematics and the aesthetics of network visualisation; 30% relate to VLSI applications, with 21% accounted for by placement and 9% by scheduling; in approximately 20% of force-directed algorithm studies, they are applied for social information visualisation; and biological network visualisation and sensor placement and localisation account for 10% and 3%, respectively. These statistics suggest that most applications of force-directed algorithms can be formulated as a problem of network visualisation, which, in turn, can be understood as problem of combinatorial optimisation — to find a visual drawing of an input network topology in a way that optimises functions of interest.

We have adopted a simple approach to classify the papers which are related to force-directed algorithms. The data sources of the papers reviewed in this survey are from ACM Digital Library [3] and Scopus [4]. The paper submission count of force-directed algorithms classified by application fields is illustrated in Figure 8. First, the papers reviewed in this survey were sorted by publication year. Next, they were categorised into corresponding application domains. The results of the classification are illustrated in Figure 1(c). According to our classification results, studies of force-directed algorithm applications in VLSI have the longest history. The first VLSI study of force-directed algorithms was published in 1965 and this research domain remains popular in 2017. Aesthetics drawing became popular around 1995 and its popularity is ongoing. By contrast, force-directed algorithms for sensor placement and localisation are relatively new research domains. The first publication in this area dates to 2004 and the publication count has increased since 2008. We also found evidence of force-directed algorithmic applications for biological network visualisation dating back to 1995, with publication counts increasing dramatically from 2003 (8 papers per year on average). Finally, studies of force-directed algorithms for social information visualisation have been popular since 2005 and, to date, offer the highest publication counts among all of the research fields.

Each of these applications relates to information visualisation broadly. Information visualisation allows users to make better sense of network relationships than by simply looking at data in tabular form. However, unsupervised visualisation cannot meet these objectives. How network topologies are drawn can significantly affect how viewers understand the network. The layout and position-assignment of visualised network nodes influence how a user perceives network relationships. Identifying visualisations that convey the appropriate information to the user is thus crucial. Filtering and pattern analysis have also been applied for force-directed algorithms to discover insightful relationships and reduce clutter. These methods are especially useful in the visualisation of social data, in which metrics associated with each node are used to understand and identify unexpected network patterns more effectively.

Force-directed algorithms face a number of challenges. Most visualisation problems are NP-hard; as such, approximation methods and heuristics are often proposed, because an almost-global optimum is sufficient for most applications. In addition, force-directed algorithms currently suffer from a number of technical drawbacks. First, they are easy to converge to a localised optima. Second, even the hardware performance has been improved; the running time of force-directed algorithms is still high when producing visualisations for large networks. Third, it is time-consuming to fine-tune the parameters of a large class of networks because the suitable parameters for a particular network class are often disadvantageous for other classes.

Several literature reviews on force-directed algorithms have been published in recent years [5-10]. In [10], Battista et al. presented an annotated bibliography of algorithms for visualisation of graphs. The algorithms reported in their review can be used to visualise various types of graphs such as trees, general graphs, planar graphs, directed graphs, etc. Force-directed algorithms for visualisation of straight-line drawings were also reported in the bibliography. In [5], Gibson et al. reviewed algorithms for force-directed layouts, dimension reduction in graph layout, and multilevel techniques for computational improvements. Gibson et al. also evaluated force-directed algorithms based on aesthetic properties of the drawings such as minimising edge crossings, achieving symmetry, and uniformity on edge nodes, etc. In [7], Tamassia et al. reviewed the algorithms for symmetric graph drawing, tree drawing, spine and radial drawings, circular drawing, rectangular drawing and force-directed drawing, etc. Tamassia et al. also summarised the algorithms and tools used in different application areas such as computer security, education, computer networks, data analytics, graph drawing and cartography, social networks and biological networks. In [11], Battista et al. reviewed the algorithms for force-directed drawing, planer orthogonal –or- straight-line drawings, non-planar drawings, etc. In their review, force-directed algorithms were categorised based on spring and electrical forces, barycenter method, forces simulating graphs theoretic distance, energy functions and magnetic fields. In addition, aesthetic properties such as edge crossings, minimisation of the area of the drawing, minimisation of the length of edges, uniform edge length, uniform bends, symmetric property were also summarised.

In [6], Kobourov summarised spring systems and electrical forces in graph drawings such as graph theoretic distances approach, stress majorisation, non-Euclidean approaches, and Lombardi spring embedders. They also considered several classical algorithms in spring embedders layouts such as force-directed algorithms, barycentric method, and multiscale methods for dynamic graphs. In [9], Brandenburg et al. compared five force-directed algorithms for drawing graphs in which the positions of the nodes are randomised. Their experiments aimed to evaluate the performance of force-directed algorithms in terms of uniformity in edge length and node distribution. In contrast to previous surveys, in this paper, we provide a comprehensive summary and full roadmap for the state of the art in force-directed algorithms in terms of latest research domains and models including social information visualisation, biological network visualisation, sensor networks, routing algorithms, scheduling, and graph drawing. An overview of the classification of existing force-directed algorithms is also provided in this survey.

In our survey, 230 papers related to force-directed algorithms have been reviewed. To find these papers, we implemented a web mining tool using Java programming language to parse search results from the ACM Digital Library [3] and Scopus [4]. We used four keywords (―force directed algorithms‖, ―force-directed algorithms‖, ―force-directed‖ and ―force directed‖) to filter relevant papers. The search results from the ACM Digital Library [3] contain the attributes such as authors, title, keywords, abstract and result highlight of papers. The search results from Scopus [4] contain similar attributes except the highlights. Moreover, we applied following filters to remove irrelevant and redundant results in order to improve the accuracy:

1. The abstract, highlights, the keywords, or the title of the paper must contain at least one of the four keywords used in the searching.
2. Papers returned from partial match were omitted. For example, They force are applied … directed … algorithm, … directed …, force …, …algorithm.

We also checked the first author and the title of the paper to remove duplicate publications. Figure 2 illustrates state-of-the-art studies and milestones in various force-directed models, including the accumulated force model, the energy function minimisation model, the combinatorial optimisation model, the multilevel model, the multidimensional scaling model and the clustered model. Our findings suggest that many papers are application studies, in which force-directed algorithms are used but without detailed formulation. Application studies usually adopt and/or revise existing force-directed algorithms to achieve the objectives of specified tasks. Because of this, our survey is divided into two parts. For those studies adopting force-directed algorithms to resolve schematic drawings and placement tasks, we first summarise them in our survey in terms of application domains and methods. We then conclude the formulation (model) of notable force-directed algorithms which have been used in application discussed in the first part.

The structure of the survey is as follows: Section 2 presents an overview about the notable force-directed algorithms that have been used most often across the different application domains. Section 3 introduces force-directed algorithms for applications in schematic drawings and placement. Section 4 concludes the survey by summarising patterns across the literature.

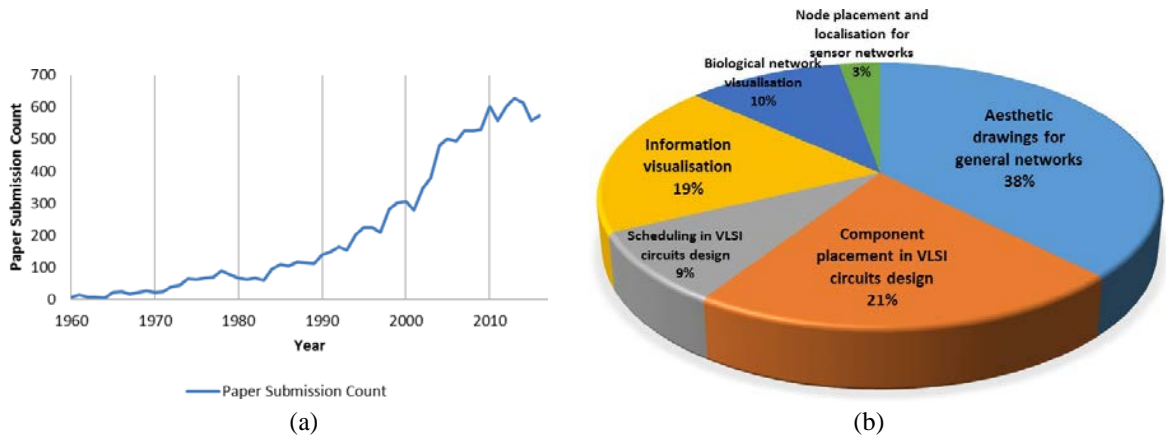

(a)                                        (b)

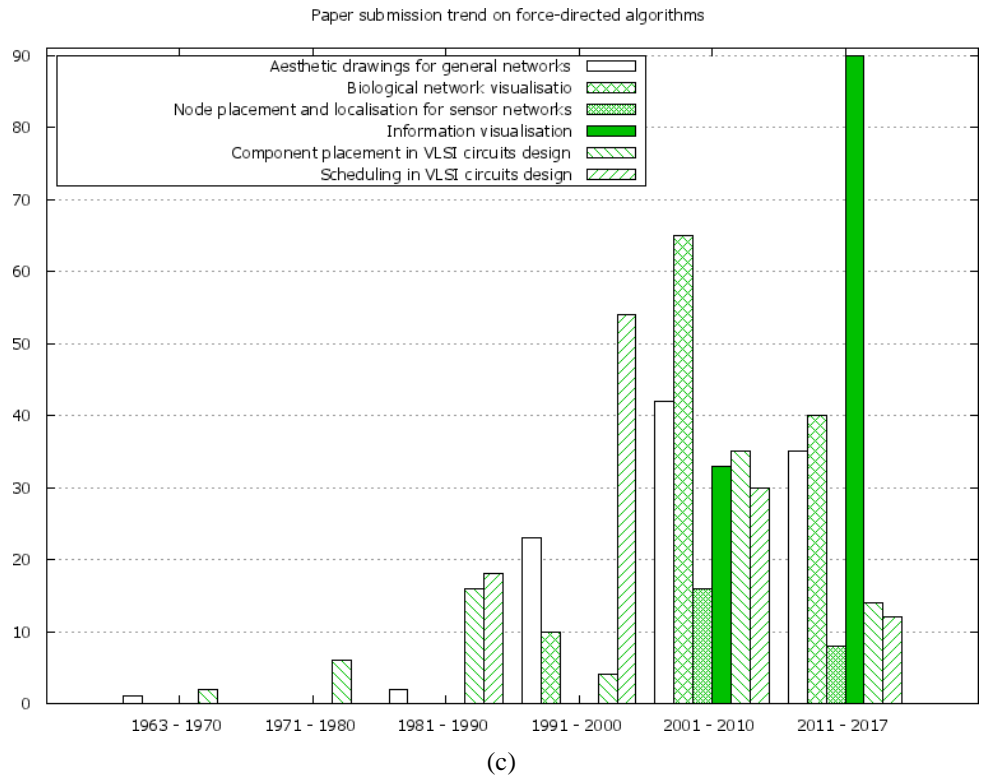

(c)

Figure 1 (a) Annual paper submission count related to force-directed algorithms; (b) Catalogues of the papers reviewed in this survey; (c) Paper submission trend on force-directed algorithms.

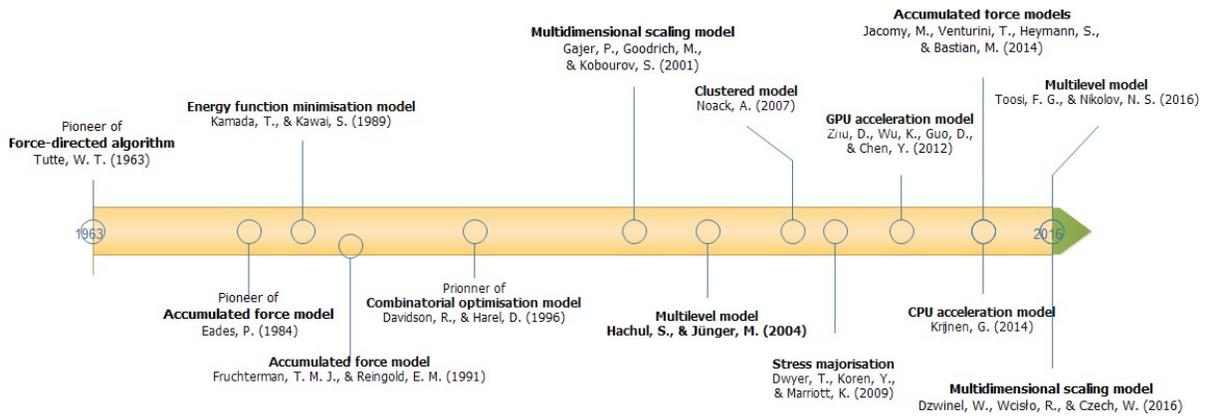

Figure 2 Studies of force-directed models.

## 1.1    NOTATIONS AND CONVENTIONS

For the purpose of the survey, the notation $G = (V, E)$ represents a network $G$, including a set of nodes $V$ and edges $E$ between these nodes. The visual drawing of a network is a picture of a network that assigns a position to each node and a curve to each edge. A connected network is a network in which for each pair $u$, $v$ of nodes, there is always a path between $u$ and $v$.

## 2 FORCE-DIRECTED ALGORITHMS

Force-directed algorithms can be divided into classical and hybrid algorithms according to their characteristics and computational modelling. The overview of force-directed algorithms is illustrated in Figure 3. Classical force-directed algorithms are usually based on physical laws, specifically in ways that simulate a spring system. Full descriptions of classical force-directed algorithms are described in section 2.1. Hybrid force-directed algorithms are designed for large and complex networks. These algorithms use heuristics to improve the performance of classical force-directed algorithms. Hardware acceleration and multilevel methods are also popular in improving the performance. Full descriptions of notable hybrid force-directed algorithms are described in section 2.2.

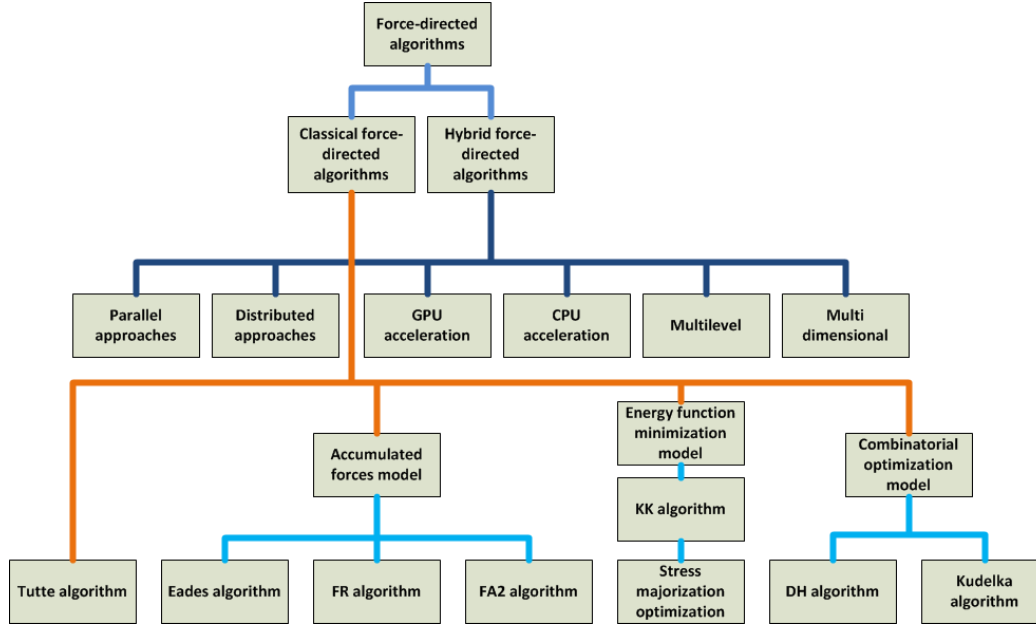

Figure 3 Overview of force-directed algorithms.

The pioneer of force-directed algorithms, the Tutte algorithm, was first proposed in 1963 [197]. The Tutte algorithm is based on the barycentric method [7, 198] and is applicable for tri-connected and planar graphs. A tri-connected graph is a connected graph such that deleting any two nodes results in a graph that is still connected. The force function of a node $v$ of the Tutte algorithm is defined as follows:

$$F(v) = \sum_{u,v \in E} (p_u - p_v) \qquad \qquad 1)$$

where $p_u$ and $p_v$ are the positions of node $u$ and $v$. Solving the linear equations from the result of partial derivatives of the force function of Tutte algorithm $F$ can obtain the updated $x$-coordinate and $y$-coordinate of nodes. These linear equations are defined as follows:

$$x_v = \frac{1}{deg(v)} \sum_{u,v \in E} x_u \qquad \qquad 2)$$

$$y_v = \frac{1}{deg(v)} \sum_{u,v \in E} y_u \qquad \qquad 3)$$

where deg(v) is the number of edges attached to node v. $x_v$, $y_v$ are the x-coordinate and y-coordinate of node v. An example of the Tutte algorithm is illustrated in Figure 4. Nodes 1, 2, 3, 4 and 5 in Figure 4 form a strictly convex polygon. The Tutte algorithm first selects a strictly convex polygon from the graph, in which all nodes on the convex polygon should have a fixed initial position. Therefore, nodes 1, 2, 3, 4 and 5 are assigned a fixed initial position. The position of remaining nodes (i.e. 6, 7, 8) then can be computed by the Tutte algorithm.

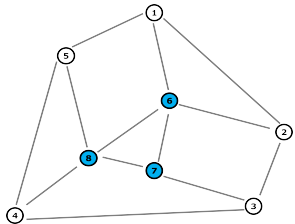

Figure 4 An example of Tutte algorithm.

## 2.1 CLASSICAL FORCE-DIRECTED ALGORITHMS

### 2.1.1 Accumulated force models

Accumulated force models follow the simulation of a spring system, in which the length of the spring is proportional to the force exerted by an extended spring. Repulsive and attractive forces are basic forces defined in the accumulated force models. Repulsive force is computed for every node pair and attractive force is computed for every adjacent node [199]. The sum of the values of repulsive and attractive forces for each node are stored in the temporary variables, which can be used for updating the nodes'positions. Most accumulated force models follow Hooke's law [200] and the footsteps of Eades' algorithm [26]. Because of this, we first introduce the principle of Eades algorithm in the section 2.1.1.1. We then introduce the successors of Eades algorithm, Fruchterman-Reingold algorithm and ForceAtlas2 algorithm, in sections 2.1.1.2 and 2.1.1.3, respectively.

#### 2.1.1.1 Eades algorithm

The idea of Eades' spring-embedded algorithm is to model a network as a magnetised system with rings representing nodes and the length of edges represented by the spring. Eades [26] was the first algorithm to consider attractive and repulsive forces. The attractive force $f_a$ is applied to nodes that have a direct connection by an edge (i.e. $(i,j) = 1$), and the repulsive force $f_r$ is applied to nodes that have an indirect connection (i.e. $(i,j) > 1$). The attractive and repulsive forces of Eades algorithm are defined as follows:

$$f_a(i,j) = C_a \log \frac{d(i,j)}{d_0} \qquad \text{4)}$$

$$f_r(i,j) = C_r \frac{1}{d(i,j)^2} \qquad \text{5)}$$

where $d(i,j)$ is the distance between node $i$ and $j$, $d_0$ is the ideal edge length, and $C_a$ and $C_r$ are the constants. The aim of the algorithm is to find zero-force locations for all nodes to reach a state of equilibrium for the spring system.

#### 2.1.1.2 Fruchterman-Reingold algorithm

The Fruchterman-Reingold (FR) algorithm [2] is based on Eades algorithm [26]. Like the Eades algorithm, the FR algorithm uses two forces, with the attractive force ($f_a$) and repulsive force ($f_r$) defined as follows:

$$f_a(d) = \frac{d^2}{k} \qquad (6)$$

$$f_r(d) = -\frac{k^2}{d} \tag{7}$$

where $d$ is the distance between two nodes and $k$ is the constant of ideal pairwise distance. For the attraction force, $f$, $k$ can be written as $a \times \sqrt{\frac{W \times H}{n}}$, and can be written as $r \times \sqrt{\frac{W \times H}{n}}$ for the repulsion force, $f_r$; where $W$ is the width of the canvas, $H$ is the height of the canvas, $n$ is the total number of nodes in the network topology, $a$ is a constant for the attraction multiplier, and $r$ is a constant for the repulsion multiplier.

The FR algorithm is executed iteratively. In each iteration, all of the nodes are moved simultaneously after the forces have been calculated. When updating the position of the nodes, the algorithm adds a ‚displacement' attribute to store the position offset of the nodes. At the start of each iteration, the initial values of the displacement for all of the nodes are calculated using the repulsion force ($f_r$). The algorithm uses the attraction force ($f_a$) to iteratively update the position of the nodes on every edge. Finally, it updates the position offset of the nodes using the displacement value.

The displacement scale, , is used as the termination condition of the FR algorithm. When the displacement scale, $s$, is lower than the threshold value, $\varepsilon$, the algorithm is terminated. When the algorithm is initialised, the value of the displacement scale, $s$, is set to $\frac{W}{10}$. This value is updated in each iteration according to the iteration count and the maximum number of iterations set by the user.

### 2.1.1.3 *ForceAtlas2 algorithm*

ForceAtlas2 (FA2) was proposed by Jacomy et al. [22] to satisfy speed and precision for network visualisation. The algorithm extends the LinLog [32] and FR algorithm [2]. Its authors proposed a revised attractive force based on the LinLog model [32] and defined as follows:

$$F_a(n_1, n_2) = log(1 + d(n_1, n_2)) \tag{8}$$

where $d$ is the distance between nodes $n_1$ and $n_2$. Moreover, a degree-dependent repulsion model was proposed in the FA2 algorithm to reduce the repulsive forces. This repulsion model increases the chances of lower-than-average-degree nodes connecting to higher-than-average-degree nodes.

$$F_r(n_1, n_2) = k \times \frac{(deg(n_1) + 1) \times (deg(n_2) + 1)}{d(n_1, n_2)} \tag{9}$$

where $k$ is a constant of ideal pairwise distance, as used in the FR algorithm [2], $d$ is the distance between nodes $n_1$ and $n_2$ and $de(n)$ is the number of edges associated with the node $n$, including in- and out-degree edges. In addition, the FA2 algorithm also uses gravitational force and strong gravitational force. Jacomy et al. [22] concluded that strong gravitational force may be useful only for specific types of networks. The definition of these gravitational and strong gravitational forces are defined as follows, respectively:

$$F_g(n) = k \times (deg(n) + 1) \tag{10}$$

$$F_{sg}(n) = k \times (deg(n) + 1) \times d(n) \tag{11}$$

### 2.1.2 **Energy function minimisation model**

In contrast to the accumulated force model, the energy function minimisation model uses the spring system to minimise the difference between the visual distance and theoretical graphed distance, and this is accomplished by solving (minimising) an energy function. They do not consider attractive and repulsive forces separately, but rather in conjunction to minimise an energy function. That is, if the visual distance of a pair of nodes is closer than their corresponding theoretical graphed distance, they repel each other; otherwise, they attract each other. The Kamada-Kawai algorithm is the pioneering algorithm for energy function minimisation models. The description of the Kamada-Kawai algorithm is given in section 2.1.2.1 and a technique to improve the energy function minimisation model is summarised in section 2.1.2.2.

### 2.1.2.1 Kamada-Kawai algorithm

In the Kamada-Kawai (KK) algorithm [31], nodes are placed so that their visual distance within the drawing is proportional to their theoretical graphed distance. As this goal cannot always be achieved for arbitrary network topologies, the key idea behind the algorithm is to use a spring model in such a way that the principle of energy function of the network topology is minimised. The energy function E is:

$$E = \sum_{i=1}^{n-1} \sum_{j=i+1}^{n} \frac{1}{2} k_{i,j} \left( \left| p_i - p_j \right| - l_{i,j} \right)^2 \tag{12}$$

where $k_{i,j}$ is the stiffness of a spring between nodes $i$ and $j$, $l_{i,j}$ is the ideal distance of a spring between nodes $i$ and $j$, and $p_i$ and $p_j$ are the visual positions of nodes $i$ and $j$, respectively. That is, the KK algorithm finds a visual position for each pair of nodes $i$ and $j$, and their Euclidean distance is proportional to $l_{i,j}$. Here, the KK algorithm defines a diameter matrix that stores theoretical graphed distances ($d_{i,j}$) of the nodes. $d_{i,j}$, which represents the hop count between nodes $i$ and $j$. $d_{i,j}$ is the shortest hop count between nodes $i$ and $j$. The ideal distance of a spring ($l_{i,j}$) between nodes $i$ and $j$ is defined as follows:

$$l_{i,j} = \frac{L_0}{\max_{i<j} d_{i,j}} \times d_{i,j} \tag{13}$$

where $L_0$ is the side length of the drawing frame and $\max_{i<j} d_{i,j}$ is the diameter of the network topology. Moreover, the stiffness of a spring between nodes $i$ and $j$ is calculated as follows:

$$k_{i,j} = \frac{K}{d_{i,j}^2} \tag{14}$$

where $K$ is a scaling and $d_{i,j}$ represents the theoretical graphed distances of nodes $i$ and $j$. The KK algorithm then seeks a visual position for every node $v$ in the network topology and tries to decrease the energy function in the whole network. That is, the KK algorithm calculates the partial derivatives for all of the nodes in the network topology in terms of every $x_v$ and $y_v$ that are zero (i.e., $\frac{\partial E}{\partial x_v} = 0$ and $\frac{\partial E}{\partial y_v} = 0, for\ 1 \leq v < n$). However, solving all of these non-linear equations simultaneously is unfeasible because they are dependent on one another. Therefore, an iterative approach can be used to solve the equation based on the Newton-Raphson method. At each iteration, the algorithm chooses a node $m$ that has the largest maximum change ($\Delta_m$). In other words, the node $m$ is moved to the new position, where it can reach a lower level of $\Delta_m$ than prior. Meanwhile, the other nodes remain fixed. The maximum change ($\Delta_m$) is calculated as follows:

$$\Delta_m = \sqrt{\left(\frac{\partial E}{\partial x_m}\right)^2 + \left(\frac{\partial E}{\partial y_m}\right)^2} \tag{15}$$

### 2.1.2.2 Stress majorisation optimisation

In force-directed algorithms such as the KK algorithm [31], visual distance is proportional to the theoretical graphed distance. Stress majorisation optimisation [201-203] is a technique to minimise energy function via majorisation. This technique improves the visual drawing of network topologies iteratively. The principle of majorisation optimisation is to construct a sequence of quadratic forms in which each iteration binds the stress function. The stress function then monotonically decreases (never increases) with every iteration. Thus, a lower value for the energy function is achieved in the same running time [203]. Unlike the KK algorithm, then, the stress function optimised via majorisation is guaranteed to converge [7]. Stress majorisation optimisation is useful for large and clustered networks, especially for applications to social information visualisation [204, 205].

### 2.1.3 Combinatorial optimisation model

Combinatorial optimisation models are probabilistic algorithms, often inspired by evolutionary mechanisms. Simulated annealing, differential evolution and genetic algorithms use a number of measures to improve a candidate solution and to optimise a problem iteratively. Although these algorithms share many similar properties, they still have distinctive features, including population determination, strategies to search the solution state space, etc. [206].

#### 2.1.3.1 Davidson-Harel algorithm

The process of simulated annealing is inspired by the physical cooling process of molten materials. Molten steel will crack and form bubbles that make it brittle if cooled too quickly. The steel must therefore be cooled evenly for a better result — a process known as annealing in metallurgy [7, 207, 208]. The Davidson-Harel (DH) algorithm [29] uses a simulation of the annealing process to prevent nodes from moving too close to non-adjacent edges and to minimise edge crossings. An energy value $E$, attraction force $f_a$ and repulsion force $f_r$ are used in the simulation. The energy value ($E$) is the sum of all attraction forces ($f_a$) and repulsion forces ($f_r$) which can be calculated as follows:

$$E = \sum_{i=1}^{n-1} \sum_{j=i+1}^{n} f_a(\sqrt{(x_i - x_j)^2 + (y_i - y_j)^2}) + f_r(\sqrt{(x_i - x_j)^2 + (y_i - y_j)^2}) \qquad (16)$$

A node $i$ is randomly selected from the network on initialisation. The DH algorithm then creates a temporary node $j$, and assigns a position to the node based on the position of node $i$. Therefore, a new energy value $E'$ can be calculated using the position of node $j$ and other nodes within the network.

$$E' = \sum_{v, i \in V, j \notin V, v \neq i} f_a(\sqrt{(x_v - x_j)^2 + (y_v - y_j)^2}) + f_r(\sqrt{(x_v - x_j)^2 + (y_v - y_j)^2}) \qquad (17)$$

Moreover, the DH algorithm obeys the rules of the Boltzmann distribution when the liquid is cooled slowly [209]. If $E' - E \leq 0$, then $E'$ is used as the energy of the next iteration, as $E'$ has lower energy value. If $E' - E > 0$, a probability equation is used to determine whether to use the new energy $E'$ in the next iteration. The probability equation is defined as follows:

$$p = e^{-\frac{(E' - E)}{k \times T}} \qquad (18)$$

where $T$ is the temperature variable and $k$ is the Boltzmann constant. If the probability $p$ is less than the threshold $\varphi$, then the new energy $E'$ is accepted; otherwise, the old energy $E$ will be used in the next iteration.

#### 2.1.3.2 Kudelka algorithm

The Kudelka algorithm [210] is a force-directed algorithm that aims to find a low-dimensional representation of the high-dimensional network. This allows the high-dimensional network to be visualised in low-dimensional (e.g. two- or three-) space. Sammon's mapping [211] and the differential evolution method are used in the Kudelka algorithm. Differential evolution is a population-based optimiser. It evolves a population of real encoded vectors in which the initial values of vectors are randomly chosen from within a predefined range. Differential evolution generates new vectors and operations using the real encoding of candidates. As a result, new vectors are perturbed and scaled from the existing vectors of the population. The objective of the Kudelka algorithm is to minimise the projection error function $E$, which is defined as follows:

$$E = \frac{1}{\sum_{i<j}^{m} d_{ij}^*} \sum_{i<i}^{m} \frac{(d_{ij}^* - d_{ij})^2}{d_{ij}^*} \qquad 19)$$

where $d'_{ij}$ is the distance between $X_i$ and $Y_j$. The distance between corresponding vector $Y_i$ and $Y_j$ in lower dimensional space is denoted as $d_{ij}$.

## 2.2 HYBRID FORCE-DIRECTED ALGORITHMS

Several studies have used heuristic techniques to improve the performance of force-directed algorithms and reduce execution time, enabling the algorithms to visualise large and complex networks in an efficient manner. For example, the multilevel technique simplifies networks through network abstraction processes. Distributed force-directed algorithms use parallel computing and hardware acceleration to reduce execution time for parsing large networks. The multidimensional scaling technique is useful for visualising networks' meaningful underlying dimensions. State-of-the-art studies of these heuristics are discussed and summarised in the following sections.

### 2.2.1 Parallel and hardware accelerated force-directed algorithms

The major principle of parallel computing is to solve a computational problem using multiple resources simultaneously [212, 213]. Generally, parallel computing involves the following steps:

1. A computational problem is first broken into smaller pieces of executable content that can be solved concurrently.
2. Each piece of executable content will be further broken down into a series of instructions for the Central Processing Unit (CPU) or Graphics Processing Unit (GPU).
3. Instructions from every piece of executable content are executed simultaneously on different CPU or GPU.
4. An overall coordination mechanism is used. When a task has completed the execution of instructions, it sends an acknowledgment to the coordinator before sending the result to the receiving task.

Most parallel computing frameworks [214-216] for force-directed algorithms are based on the accumulated force model. For example, the GPU parallel computing framework [217] was proposed for identifying the k-nearest neighbours, the results of which were then utilised to speed up the FR algorithm [2]. A distributed force-directed algorithm in an open source distributed computing framework called Giraph1 [218] was implemented in Amazon's cloud computing infrastructure PaaS (Platform as a Service) [219]. Arleo et al. [218] claimed that the algorithm can process networks with up to million edges. A parallel FR algorithm [2] based on Open Computing Language (OpenCL) was proposed by Krijnen [220] and Wang et al. [221]. OpenCL programs can be executed across heterogeneous platforms with modern CPUs, GPUs, and microprocessor designs [222]. There are also parallel force-directed algorithms [223-225] based on the Message Passing Interface (MPI). MPI is defined by a group of parallel computing vendors and applications specialists[2] as a specification for a standard library for message passing in distributed computing.

### 2.2.2 Multilevel force-directed algorithms

The multilevel technique for force-directed algorithms involves concepts from network abstraction and can be divided into two main phases. In the first phase, called ‗coarsening', the original network is split into a sequence of coarse networks with decreasing sizes. This simplifies the combinatorial structure of the network by selecting the coalescent pairs of adjacent nodes to construct a new network. The selection process is repeated recursively to abstract a sequence of such coarse networks. The process of energy optimisation (minimisation) is then performed across these coarse networks such that they are optimised using the global properties from the original network. The second phase is called refinement and involves successive drawings of fine networks computed from the smallest coarse networks. Finer networks are optimised using the locally determined properties from the related coarse network. As a result, it can decrease running time because the energy minimisation process considers only a small amount of neighbourhoods at once [226]. Many studies have proposed using the multilevel technique for force-directed algorithms [227-230]. There are also studies that extend the multilevel technique to the classical force-directed algorithms such as multilevel KK algorithm [30] and multilevel FR algorithm [24].

### 2.2.3 Force-directed algorithms with multidimensional scaling

High-dimensional data usually have a large number of variables instead of a large number of duplicated records. The multidimensional scaling technique is widely used in force-directed algorithms for high-dimensional data reduction. The objective of the multidimensional scaling technique is to find meaningful underlying dimensions so that observed similarities and dissimilarities from the investigated networks can be discerned easily. The principle behind multidimensional scaling was developed by Torgerson [231] which uses the distance of edges as a metric. Nodes are projected into a smaller space that satisfies the constraint of the metric (the distance of edges). Many studies have adopted multidimensional scaling for force-directed algorithms to visualise high-dimensional data in which the distances between pairs of data are preserved [201, 232-239]. Multidimensional scaling is also useful for energy function minimisation modelling, as it can improve the layout of networks with high-degree nodes. Dwyer et al. [240] and Dzwinel et al. [241] proposed a multidimensional scaling KK algorithm [31] with the use of stress majorisation optimisation. The energy function proposed by Dzwinel et al. [241] is defined as follows:

$$E = k_{nn} \sum_i^N ( \sum_{j \in O_{nn}(i)}^{nn} d_{ij}^{n\,2} + c \times \sum_{k \in O_{rn}(i)}^{rn} (1 - d_{ik}^n)^2) \qquad 20)$$

where $k_{nn}$ and c are constants and configured by users. $d_{ij}^n$ is the distance of node $i$ and $j$ in the visual drawing. $O_n(i)$ is the nearest neighbourhood of node $i$ (i.e. hop count equals to 1). $O_r(i)$ is the random neighbourhood of node $i$ (i.e. hop count greater than 1).

## 3 APPLICATIONS OF FORCE-DIRECTED ALGORITHMS

This section reviews five categories of application domains in which force-directed algorithms have been adopted: (a) aesthetic drawings for general networks, (b) component placement and scheduling in high-level synthesis of very-large scale integration (VLSI) circuits design, (c) information visualisation, (d) biological network visualisation, and (e) node placement and localisation in sensor networks.

## 3.1 FORCE-DIRECTED ALGORITHMS IN AESTHETIC DRAWINGS FOR GENERAL NETWORKS

Force-directed algorithms can be used to produce schematic drawings from network topology alone, even without additional information about its nodes and edges. However, many applications of force-directed algorithms involve an implicit aesthetic problem in how to schematise topological renderings. The importance of such schematics is that its depiction can significantly influence how the topology is understood. For example, what are the aesthetic properties of the most coherent schematics? How can the aesthetic quality of schematics be measured? To understand these questions, we need to clarify the characteristics and objectives of a schematic. The fundamental factor is its layout. For example, in the polyline drawing (see Figure 5 (a)), each edge is a polygonal chain. Whereas in the straight-line drawing (see Figure 5 (b)), each edge is a straight-line segment. In the orthogonal drawing (see Figure 5 (c)) [8], each edge represents a horizontal and vertical segment. Numerous visualisation tools have been implemented for visualising networks in different layouts, most developed for straight-line drawing, such as GraphED3 [12], COMAIDE [13], LayoutShow [14], Graphael [15] and OpenOrd [16]. A visualisation tool based on orthogonal drawing is also proposed in [17].

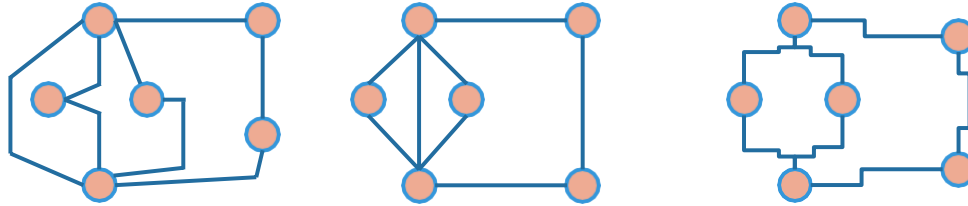

Figure 5 (a) Polyline drawing   (b) Straight-line drawing   (c) Orthogonal drawing

Creating aesthetically appealing schematics has the practical aim of revealing a structure's pattern, rather than being merely a quest for the beautiful [18]. Therefore, researchers have defined the properties of a schematic based on its fundamental factors. Force-directed algorithms can be used to produce schematics that adhere to the properties of aesthetic drawing [9, 19, 20]. The properties of aesthetic drawing include : 1) edge lengths should be uniform; 2) the number of edge crossings should be minimised; 3) the size of crossing angles should be uniform; 4) the crossing angle should be minimised; 5) the standard deviation of edge length should be low; 6) the angle formed by any two neighbouring edges should be minimised; 7) the number of bends in polyline edges should be minimised; 8) nodes and edges should be affixed to an orthogonal drawing; and 9) the network should be represented as symmetrically as possible. In [21], Tunkelang proposed a force-directed approach for drawing undirected graphs. It is based on the accumulated force model that includes repulsive and attractive forces. Repulsive forces are computed between any two nodes and attractive forces are calculated between two adjacent nodes. Repulsion among nodes are used to avoid situations where nodes are placed too close to each other. Attraction forces are used to prevent nodes from being too far away from each other. According to the principles of the accumulated force model [22], nodes pull far away from each other if they are not adjacent. Besides, the model tries to maintain uniform edge lengths among adjacent nodes to minimise edge crossings. The repulsive and attractive forces of the proposed algorithm are defined as follows:

$$f_r(d) = \frac{w_r}{d^2}$$
(21)

$$f_a(d) = w_a d$$
(22)

where $d$ is the length of edge and $w_r$ and $w_a$ are constants. The objective of the algorithm is to find an optimal value $d$ so that the sum of attractive and repulsive forces (i.e. $f(d) + f_a(d)$) is minimal. In addition, a force-directed algorithm was also proposed to produce schematics based on the fitness function of a genetic algorithm (GA) [23]. A number of studies have adopted similar approaches in the literature. Due to the page limit, we summarise them in terms of the models used and the property of the aesthetic drawing in Table 1.

Table 1 Forced-directed algorithms for aesthetic visual drawings.

| Catalogue | Property of aesthetic drawing | Adopted by proposed force-directed algorithms | Models used in force-directed algorithms |
|---|---|---|---|
| Node | Distribute nodes evenly | [2], [22], [24], | Accumulated force model |
| | | [25], [26], [27], [28] | |
| | | [29] | Combinatorial optimisation model |
| | | [30] | Energy function minimisation model with a multiscale approach |
| | | [31] | Energy function minimisation model |
| | | [23] | Energy function minimisation model with a fitness function in GA |
| | Cluster similar nodes | [32] | Energy function minimisation model |
| | Nodes should not overlap | [33] | Accumulated force model |
| | Nodes that are not adjacent should be far away from each other | [21] | Accumulated force model |
| | | [29] | Combinatorial optimisation model |
| Edge | Minimise edge crossings | [2], [21], [22], [24], [26], [27], [28], [34] | Accumulated force model |
| | | [29] | Combinatorial optimisation model |
| | | [35] | Multilevel force-directed algorithm |
| | | [36], [37] | Energy function minimisation model |
| | Minimise edge bends | [35] | Multilevel force-directed algorithm |
| | Keep edge lengths uniform | [2], [21], [22], [24] | Accumulated force model |
| | | [31] | Energy function minimisation model |
| | | [23] | Energy function minimisation model with an additional fitness function of genetic algorithm |
| | | [30], | Energy function minimisation model with a multiscale approach |
| | Minimise edge length | [28] | Accumulated force model |
| | | [23] | Energy function minimisation model with an additional fitness function of genetic algorithm |
| all layo | Display of symmetries | [25], [26] | Accumulated force model |

| | | [31], [37] | Energy function minimisation model |
|---|---|---|---|
| | | [30] | Energy function minimisation model with a multiscale approach |
| | | [23] | Energy function minimisation model with an additional fitness function of genetic algorithm |
| | Maximise the angles among incident edges | [28], [38] | Accumulated force model |
| | | [35] | Multilevel force-directed algorithm |
| | | [36] | Energy function minimisation model |
| | | [23] | Energy function minimisation model with an additional fitness function of genetic algorithm |
| | | [39] | |
| | The angles between edges incident on the same node should be as uniform as possible | [28], [39] | Accumulated force model |
| | | [36] | Energy function minimisation model |
| | | [23] | Energy function minimisation model with an additional fitness function of genetic algorithm |
| | Orthogonality | [17], [40] | Accumulated force model with an additional octilinear magnetic force [41] for orthogonal drawing |
| | Minimise the size of visual drawing | [37] | Energy function minimisation model |

## 3.2 FORCE-DIRECTED ALGORITHM IN COMPONENT PLACEMENT AND SCHEDULING IN VLSI CIRCUITS DESIGN

Technical Terms such as ‚module', ‚cell', ‚pin' and ‚component' are widely used in the studies of very-large-scale integration (VLSI) circuits. They are similar to the concept of nodes in graph theory. To make the terms consistent in this survey, we use the term ‚node'. Force-directed placement algorithms and force-directed scheduling are widely used in the design and manufacturing for VLSI circuits. An example of components from a VLSI circuit board is illustrated in Figure 6 (a) (generated by the visual5602 simulator [42]). The roadmap and approaches for these techniques are discussed in the following subsections.

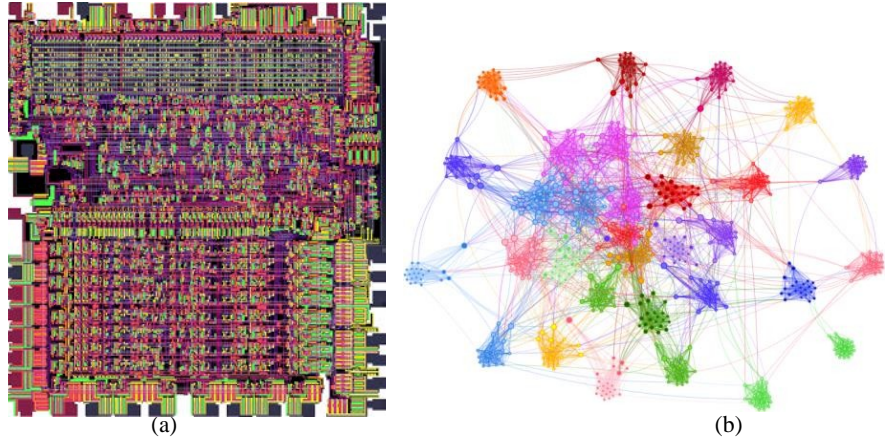

Figure 6 Visualisations of (a) components of a VLSI circuit (b) a clustered network.

### 3.2.1 Force-directed placement algorithms

The nodes in VLSI circuits can be integrated circuits, transistors, resistors and capacitors. The interconnection topology of the VLSI circuits is known. The objective of force-directed placement algorithms in this context is to determine the optimal location of every node with respect to every other node such that the length of edges in the interconnection topology is minimised [43]. Force-directed placement algorithms can obtain fairly non-overlapping placements on circuit boards without the use of additional means of optimisation [44] and, as such, have proven popular in applications to VLSI circuit boards since the 1960s [45-51].

#### 3.2.1.1 Pioneer approaches

Fisk and Isett [45] pioneered a system called ACCEL using two forces (i.e. attractive and repulsive forces) for the placement of nodes. Urban et al. [47] proposed a system called SHARPCLAW using similar forces [45]. Quinn and Breuer [46] and Quinn Jr [43] proposed similar systems based on Hooke's Law, with repulsive and attractive defined as follows:

$$F_r(u,v) = -\frac{K_r}{\sqrt{\dfrac{(x_u - x_v)^2}{(w_u - w_v)^2} + \dfrac{(y_u - y_v)^2}{(ℎ_u - ℎ_v)^2}}} \tag{23}$$

$$F_a(u,v) = -K_a \sqrt{\frac{(x_u - x_v)^2}{(w_u - w_v)^2} + \frac{(y_u - y_v)^2}{(ℎ_u - ℎ_v)^2}} \tag{24}$$

where $K_r$ and $K_a$ are the constants for repulsive and attraction forces, $x_u$, $y_u$ are the $x$-coordinate and $y$-coordinate of the node $u$. $w_u$ and $ℎ_u$ are the width and height of the node $u$.

#### 3.2.1.2 Modern approaches

Numerous notable force-directed placement algorithms and open-source systems have been developed since the 1990s. Most are based on solving a quadratic cost function to optimise node placement and achieve minimal edge lengths on the circuit board [52]. Force-directed relaxation methods are often used to solve the quadratic cost function. Force-directed relaxation is an iterative method in which nodes are either assigned random or fixed locations on initiation. One node is then selected at each iteration and moved to a target point determined by the forces or cost functions defined in the force-directed placement algorithms [53]. Popular algorithms include Kraftwerk [54], Kraftwerk2 [55], FAR [56], mFAR [57], FDP [58, 59], FastPlace [60], FastPlace 3.0 [61], RQL [62], SimPL [63], etc., the objective of which is to evenly distribute electromechanical components (nodes) on the circuit board, minimise the wire (edge) length and produce an overlap-free layout [64]. For example, the Kraftwerk [54] algorithm formulates the quadratic cost function $F$ defined as follows:

$$F = \sum_{u,v \in V} \frac{1}{2}(w_{uv,x} \times (x_u - x_v)^2 + w_{uv,y} \times (y_u - y_v)^2)$$

(25)

where $x_u$ and $y_u$ are $x$-coordinate and $y$-coordinate of node $u$. $w_{uv,}$ is the weight of edge $uv$ on $x$-axis (horizontal), $w_{uv,x}$ is the weight of edge uv on $y$-axis (vertical). The weight used in the $x$-axis and $y$-axis from equation (25) is different because the node (electromechanical component) placed on the VLSI circuit board is quadrilateral. We also found several extensions of the Kraftwerk algorithm [54] proposed for application in VLSI circuits [65-68]. A similar algorithm called FastPlace was proposed by Viswanathan and Chu [60]. FastPlace is also based on a force-directed relaxation precept that aims to evenly distribute nodes on the circuit board. This can be done by minimising the cost function, which is similar to equation (25). In contrast to others, Viswanathan and Chu [60] applied a post-processing technique called ‗cell shifting' to reallocate the positions of nodes that overlap as a result of force-directed placement. Pan et al. [61] also proposed an improved extension of the FastPlace algorithm, called FastPlace 3.0, which adopts a multilevel technique and uses congestion constraints [69] to place nodes evenly.

### 3.2.1.3    Partitioning and clustering based approaches

Goto [70] used a force-directed placement algorithm to divide nodes on the circuit board into two parts: an initial placement and an iterative improvement [71]. Nodes have pre-assigned (fixed) positions in the initial placement, and the force-directed algorithm calculates node locations during the improvement phase only. An algorithm based on [70] was proposed by Chang [72]. The objective of the algorithm is to find optimal regions on the circuit board to place nodes. The algorithm extends the median formulation proposed by [70] which identifies optimal regions and then applies a force-directed algorithm to calculate nodal positions within each optimal region. A force-directed placement algorithm based on clustering was also proposed by Odawara et al. [73] in which ‗seed elements‗, such as CPU and ROM from the circuit board, are first identified. Nodes close to seed elements are then grouped together to construct clusters. Finally, the relative position of each cluster is calculated by the force-directed algorithm. A similar system adopted a clustering technique was suggested by Alupoaei and Katkoori [74]. In Alupoaei‗s algorithm, clique partitioning heuristics [75] were used to cluster nodes and a force-directed algorithm based on Hooke‗s Law [46] was used to determine node placement and to minimise edge lengths on the circuit board. In [76], Vorwerk and Kennings [76] introduced a multilevel clustering algorithm to extend the algorithm proposed by [59]. The Hybird First Choice [77] clustering method was used in the Vorwerk and Kennings‗s algorithm in order to improve node placement.

### 3.2.1.4    Fixed-points and pseudo edges additional approaches

The placement of standard cells is another major application in VLSI circuits. Standard cells function as nodes with standard heights but varying widths. Numerous studies focus on the placement of standard cells. For example, some have used the cost function from the Kraftwerk algorithm [54] to determine the placement of standard cells [65]. Chou and Lin [78] located standard cells by adding additional pseudo-edges on the circuit board. In this algorithm, critical paths on the circuit board are first identified. Pseudo-edges will then attach to nodes that are close to critical paths to pull the position of nodes closer to the critical paths. All pseudo-edges are removed when the placement is completed. In addition, Hu and Marek-Sadowska [56] introduced an algorithm called FAR to add additional fixed-points (nodes). A fixed-point is a pseudo-node connected to a real node on a circuit board. Three types of fixed points are defined by [56]: controlling fixed points are used to keep the placement of a node unchanged, perturbing fixed points are used to disturb the current placement, and constraining fixed points are used to restrict the movement of a node. In another example, a flat force-directed placement algorithm called SimPL was proposed by Kim et al. [63] that does not rely on clustering. SimPL has a range of variants [79-82], all of which adopt a top-down geometric partitioning method called a look-ahead legaliser [83] to remove nodal overlap. SimPL's variants add fixed-points and pseudo-edges to produce even nodal distributions, for which the concept of fixed-points and pseudo-edges are adopted from the FAR algorithm [56]. In addition, a multilevel force-directed placement algorithm based on the energy function minimisation model [65] and fixed-point addition [56] was proposed by Hu and Marek-Sadowska [57].

### 3.2.1.5 Heuristic and application domain dependent approaches

Forbes [84] proposed a heuristic approach to accelerate the force-directed placement algorithm proposed by Fisk and Isett [45]. The objective of the heuristics is to reduce the total number of iterations of the force-directed placement algorithms. The movements of nodes during previous iterations are used to predict the position of a node in one or more future iterations. Spindler et al. [55] proposed an extension called Kraftwerk2, which is based on previous work [85]. The objective of the Kraftwerk2 algorithm is to balance the density of nodes and reduce and/or prevent any unused area (free space) of circuit board (i.e. save the space of circuit board). Two types of nodes are defined in the Kraftwerk2 algorithm. One has a fixed initial position (i.e. FN) and the other does not (i.e. MN). Only positions of the MN need to be determined in the Kraftwerk2 algorithm. Moreover, three forces are defined in the algorithm: Net Force $F_V^{net}$, Move Force $F_{V,u}^{move}$ and Hold Force $F_V^{hold}$. The force equation of the Kraftwerk2 algorithm is the sum of the three forces and defined as follows:

$$F = F_V^{net} + F_{V,u}^{move} + F_V^{hold} \qquad (26)$$

The forces of the Kraftwerk2 algorithm use concepts from a generic supply and demand system [86]. Spindler et al. [55] stated that the Net Force $F_V^{net}$ is used to minimise edge length. However, nodes will overlap when the edge length is too short. Therefore, Move Force $F_{V,u}^{move}$ and Hold Force $F_V^{hold}$ are added to the Kraftwerk2 algorithm to compensate the Net Force $F_V^{net}$ as a way to reduce nodal overlap. Interested readers can refer to Nam and Cong [86] for detailed definitions and explanations about the generic supply and demand system.

Heuristic approaches were also used for the placement of standard cells. A heuristic force-directed algorithm for the placement of standard cells was proposed by Hur et al. [87]. Congestion removal heuristics [69] are applied in Hur et al.'s algorithm to remove nodal overlaps. Additionally, a force-directed placement algorithm for determining the location of standard cells in 3D ICs (integrated circuits) away from high-temperature areas was also introduced [88].

Floor-planning is an application in VLSI closely related to placement. The goal of floor-planning algorithms [89, 90] is to develop a placement plan to decide topological proximity and the appropriate shapes and orientations of each block. A placement algorithm using the maze searching technique [91] was proposed by Mo et al. [92]. The algorithm was designed to minimise edge lengths on a circuit board. The maze searching technique is able to find the shortest path from a given node to another given node. The approach proposed by Mo et al. applies force-directed algorithms to the placement of nodes first and then uses the maze searching technique to re-route paths (edges) on the circuit board and minimise edge lengths.

Minimising the timing delay of circuits is another important task for VLSI. Force-directed placement algorithms based on Kraftwerk [54] proposed by Rajagopal et al. [93] aim to optimise the edge lengths and minimise the timing delay on the circuit board. A similar approach was proposed by Saxena and Halpin [94] to optimise the timing delay of circuits, which improves the repeater insertion technique [95] by using a force-directed approach based on Kraftwerk [54]. Repeater insertion techniques can reduce the time delay associated with long wire lines in circuit. In addition, Goplen et al. [96] proposed an algorithm to reduce repetitions during placement in which weightings [59] are used to reduce the repeater count. In [96], the cost function is adopted from Goplen's algorithm [65].

Besides timing delay minimisations, density information can also be used to improve force-directed placement algorithms [62, 97]. For example, improved versions of cell-shifting techniques were proposed by Viswanathan et al. [62]. These techniques adopted a Density-Aware Module Spreading algorithm [98] and extended the cost function of quadratic optimisation from [65] to improve the placement of nodes on circuit boards. Viswanathan et al. [62] used the density information to prevent nodes from being placed on areas that already contain high densities of edges and nodes.

Mixed-size integrated circuit (IC) design, in which the network contains a large number of nodes and macros, is also widely used in VLSI. In most cases, the magnitude (size) of macro force is larger than the size of nodes [99]. For this reason, placement algorithms should use smoothing approaches to place both nodes and macros on the chip areas simultaneously. A force-directed placement algorithm called FDP was proposed for the placement of mixed-size integrated circuits [58, 59]. The algorithm uses a dynamic weighting [100] of spreading forces. The cost function of FDP is defined as follows:

$$F = \sum_{u,v \in V} \frac{a_{uv}}{|p_u^{i-1} - p_v^{i-1}|} (p_u^i - p_v^i)^2 \tag{27}$$

where $a_{uv}$ represents the weight of the edges connecting node $u$ and $v$. $p_u^i$ and $p_u^{i-1}$ are the position of node $u$ at iteration $i$ and $i-1$, respectively. The objective of FDP algorithms is to minimise the cost function in equation (27).

Placement algorithms for 3D Field Programming Gate Array (FPGA) [101] consisting of multiple two-dimensional layers have become popular in recent studies. A low temperature simulated annealing method [102] can be used to determinate the final 3D layer from the two-dimensional layers. The latest 3D FPGA applications can be found in force-directed algorithms, such as those using the force-directed placement algorithm to minimise the edge lengths on each two-dimensional layer [103]. Integrating optical devices into the electronic communication system NoC (Networks-on-Chip) [104] is one example. The PLATON algorithm is proposed by [105] to place overlap-free Photonic Switching Elements (PSEs) on the circuit board. PSEs are components used in optical networking.

### 3.2.2 Force-directed scheduling algorithms

Force-directed scheduling algorithms are useful in High Level VLSI Synthesis systems [106-109]. An algorithm's description of a design behaviour can be interpreted by high-level synthesis [108]. For example, the context of encoding algorithms can be interpreted by high-level synthesis such that the hardware encoder/decoder algorithm can be implemented on integrated chips. Force-directed scheduling algorithms schedule instructions and operations for high-level synthesis to optimise the distribution of operations and reduce resource expenditure.

The initial force-directed scheduling algorithm was first proposed by Paulin and Knight [110] and, like other force-directed algorithms, it obeys Hooke's Law in physics. Paulin et al.'s algorithm attempts to balance the distribution of operations by decreasing concurrency of operations that make use of the same hardware resources. In the initial version of force-directed scheduling for the behavioural synthesis, proposed in Paulin and Knight [111], operations are divided into a number of steps, all of which aim at reducing the number of data buses, storage units and functional units while maintaining the concurrent operations assigned to them without lengthening the total execution time. Paulin and Knight [112] presented a force-directed scheduling algorithm to minimise interconnected costs of register allocation in high-level synthesis. Variants and extensions based on this pioneering work have been developed and reported in [113-122]. Classical scheduling has been used to minimise resources by finding a feasible schedule $r$ that minimises the resource costs. The schedule of classical scheduling is defined as follows:

$$f(r) = \sum_{r \in R} w_r \max_{t \in T} N_r(r, t) \tag{28}$$

where $R$ is a set of resource types in which $r \in R$. $w_r$ is the cost of a resource type $r$ and $t$ is the span of time required of a schedule $r$. However, solving equation (28) is a NP-complete problem. Therefore, Verhaegh et al.

[123] presented an iterative approach for the forced-directed scheduling algorithm used in PHIDEO [124] silicon compilers. The cost function of their iterative approach is defined as follows:

$$f(r) = \sum_{r \in R} w_r u_r + \sum_{r \in R} w_r {}^{max}_{t \in T}(N_r(r,t) - u_r) \tag{29}$$

$$u_r = \frac{1}{m}\rfloor * i \in O|r \underset{i}{\in} r + | \tag{30}$$

where $u_r$ is a constant based on the average number of operations for resource type $r$ over a schedule in which $m$ is the time span on a given schedule. $w_r$ is the cost for a resource type $r$. $N(r,t)$ is the number of operations of resource type r scheduled at time t in schedule $r$. The objective is to minimise the cost function $(r)$. Verhaegh et al. [125] also presented an iterative force-directed scheduling algorithm which reduces the time span of an entire operation schedule, as used in the silicon compiler PHIDEO [124].

Behavioural synthesis systems are generally designed for single tasks. Lee et al. [114] proposed a heuristic force-directed scheduling algorithm for multi-thread, real-time and multi-tasking synthesis systems. Lee et al.'s algorithm is based on the $A^*$ search technique and the force-directed scheduling algorithm proposed by Paulin et al. [111]. Multi-tasking synthesis systems contain a set of $k$ processors and a set of $n$ periodic real-time operations. The principle is to assign each operation to one of the processors in such a way that all operations can be scheduled within their time constraints. Lee et al.'s algorithm used the $A^*$ search technique to select processors that minimise the cost and satisfy timing constraints. Moreover, Abdel-Kader [115] used a force-directed scheduling algorithm derived from [111] to optimise loop scheduling in high-level synthesis. Loop scheduling is designed for repetitively performing a set of operations that functions similar to a loop in programming. Some extensions of [111] work were also proposed for reconfigurable architectures. For example, a force-directed scheduling for schedule operations in NATURE [126] was proposed by Zhang et al. [116]. NATURE is a hybrid nano/CMOS reconfigurable architecture. Force-directed scheduling algorithms are also useful for Dynamic Reconfigurable FPGAs (DRFPGAs) [117], owing to overlaps in the logic of DRFPGAs as time-multiplexed. Because of this, DRFPGAs need to be partitioned into multiple sub-circuit boards, thus possibly resulting in different execution times because sub-circuit boards are executed in parallel. Force-directed scheduling algorithms can be used to partition sequential circuits to optimise feasible partitions that reduce the logic and communication component costs while maintaining maximal throughput.

Force-directed scheduling algorithms for power optimisation problems in VLSI high-level synthesis systems have been popular since 2000. These algorithms are based, again, on the work of [111]. For example, some have used a force-directed scheduling algorithm to optimise power consumption while adhering to the resource and latency constraints in a behavioural synthesis system [119]. Gupta and Katkoori [120] also used a force-directed scheduling algorithm to optimise power consumption at the behavioural synthesis system. They reduced the overall dynamic power by reducing switched capacitance component usage during VLSI circuit design. Moreover, Allam and Ramanujam [121] proposed a force-directed scheduling algorithm for power optimisation that minimises the peak and average consumption. This can be done by assigning the smallest possible input voltage to every operation in a way that minimises power consumption.

Advanced driver assistance functions for intelligent automotive systems, such as predictive break assistants, adaptive cruise control and adaptive lane assistance are designed for processing sensor data. Schönwald et al. [122] proposed a force-directed scheduling algorithm for advanced drivers to map processes to processor cores with time and resource constraints. The objective of this work is to reduce the communication latency and increase the throughput to process sensor data. Similarly, Schönwald et al. [127] proposed a force-directed scheduling algorithm to consider shared memory architectures during the mapping of software processes on multiprocessor system-on-chip (MPSoC) cores. Schönwald et al. [127] suggested the use of smFDM (shared memory aware force-directed mapping) to determine the placement of processor cores.

Therefore, communication conflicts and memory access conflicts are reduced or even avoided. Force-directed scheduling algorithms were also proposed by Omnés et al. [118] to schedule real-time tasks to be executed on embedded multimedia systems. The work in Sethuraman and Vemuri [128] used a force-directed scheduling algorithm to optimise bandwidth in NoC (Networks-on-Chip) architecture by scheduling an optimal size (dimension mesh) of the network circuit.

### 3.3 FORCE-DIRECTED ALGORITHMS IN INFORMATION VISUALISATION

The primary objective of network information visualisation is to explore hidden patterns in networks and to visualise them in a simple manner. Information networks can be social networks, human relation networks, networks of business workflow, transportation maps, etc. An example visualisation of a clustered network is illustrated in Figure 6 (b) which is generated by using the vis.js visualisation tool [129]. The application of information visualisation is wide and complex. It is therefore impossible to visualise networks in orthogonal form or in a planar visual drawing in all cases. Moreover, certain information networks' nodes and edges may contain additional properties (attributes). These properties do not exist in general networks. Because of this, applications of information visualisation may use variant layouts to present the data. For example, metro map diagramming is useful for visualising the transportation map as a schematic [40]. With fisheye views [130, 131], the network representation can enlarge regions located near specified nodes while contracting distant regions by varying edge length. However, while enlarging a special region may be useful in two-dimensional planes, it may not be applicable for high-dimensional data. Parallel coordinate diagramming can be used to project high-dimensional data onto two dimensions [132]. Parallel coordinate diagrams draw $n$ vertical lines equally spaced to represent the n-dimensional space. Corresponding nodes are drawn on the dimensional space (vertical line) and the line represents the relation between a pair of nodes [133]. Lombardi-Style diagrams [134] are useful for information visualisation in which the edges of the visual drawing are curvilinear [135].

Even with these techniques, the amount of complexity makes visual interpretability for humans difficult. Chae [136] suggested visualising large networks on a tiled monitor wall, in which monitors are placed next to each other and data distribution related to their corresponding display nodes are only displayed. Edge crossing deduction is also crucial for visualising information about large networks, as this makes the representation appear cluttered and ugly. One example is the 1/4-SHPED (i.e. Symmetric Homogeneous Partial Edge Drawing), as proposed by Bruckdorfer et al. [137]. Nodes of 1/4-SHPED are represented as points and edges as two pieces (also called stubs) of a straight-line segment, each adjacent to a node, without any edge crossings, and with stub size 1/4 of the total edge length. Edge bundling is another technique which group edges into bundles to decrease the density of lines for reducing clutter [131-133]. Moreover, Debiasi et al. [33] proposed an accumulated force model to visualise the network by geographical flow map [33] as a way to prevent edge crossings. Each flow consists of start, intermediate and target nodes and three forces are defined in the proposed algorithm: electrostatic (attractive) force, stress force and rejected (repulsive) force. The force equation of node $v$ is the sum of the three forces and can be defined as follows:

$$F(v) = \sum_{s \in S} F_e(v, s) + \sum_{t \in T} F_r(v, t) + F_s(v) \tag{31}$$

where $S$ is the intermediate nodes interacting with node $v$. $T$ is the nodes near to the node $v$. The purpose of electrostatic force ($F_e$) is similar to attractive force, which is defined as follows:

$$F_e(v, s) = \frac{1}{\|v - s\|} \times \hat{v} - s \tag{32}$$

where $\hat{v} - s$ is the unit vector of node $v$ and $s$. $\|v - s\|$ is the norm of node $v$ and $s$. Stress force enables the node to move towards the flow with higher magnitude. The definition of stress force is as follows:

$$F_s(v) = (v_{i-1} - v) + (v - v_{i+1}) \tag{33}$$

where $v_{i+1}$ is the ancestor node of $v$ and $v_{i+1}$ is the child node of $v$. Rejected forces are used to avoid any overlapping between intermediate and target nodes, and are defined as follows:

$$F_r(v,t) = -1 \times F_e(v,s) \tag{34}$$

There are approaches which utilize forces and reduce clutter for graph visualisation in which the size of nodes in the graph is the variant. Cui et al. [134] used a force-directed model to visualise word clouds in which the size of a word (i.e. each word represents a node) is determined by the word frequency in the time slot. Moreover, Gu et al. [135] adopted the FR algorithm to visual large texts and image datasets on a large video wall. We summarise the relevant studies on information visualisation, their objectives, and corresponding force-directed algorithms used to visualise the networks in Table 2.

Table 2 Information visualisation studies that adopted force-directed algorithms.

| Force-directed algorithm adopted by the study | Study | Objective of the study |
|---|---|---|
| Eades's algorithm | [1] | – To visualise networks using tree-structured hierarchies<br>– Increase the readability of a network |
| | [2] | – To make edges conform to particular orientations |
| | [3] | – To transform the Extensible Stylesheet Language Transformations (XSLT) document to network layout. XSLT is a language for transforming vector images and documents in the XML encoding |
| | [4] | – To visualise networks with non-uniform nodes (i.e. the size and shape of nodes are variant) |
| | [5] | – To visualise the relation of people in online social networks |
| | [6] | – To visualise networks using grid layouts |
| | [7] | – To visualise web traffic |
| FR algorithm | [8] | – To visualise networks in which nodes have nontrivial sizes |
| | [9] | – To produce visual drawings of hypergraphs<br>– Hypergraphs can be viewed as an extension of classical networks in which an edge can join any number of vertices |
| | [10] | – To visualise networks with non-uniform nodes (i.e. the size and shape of nodes are variant) |
| | [11] | – To visualise exploration of network traffic over time |
| | [12] | – To visualise email networks |
| | [13] | – To visualise transportation networks |
| | [14] | – To visualise networks in which edges are curvilinear (Bézier curve) |
| | [15] | – To visualise weighted networks in which each edge is associated with a real number representing its importance |
| | [16] | – To assess homophily [17] in networks |
| | [18] | – To visualise the actors holding neutral opinion polarities |
| | [19] | – To visualise the volume of movement in flow maps |
| KK algorithm | [20] | – To show proximity between nodes such that their distances in the visualisation reflect distances in the network<br>– topology |
| | [21] | – To visualise networks in which edges are curvilinear |
| | [22] | – To visualise the structure of ER diagram |
| | [23] | – To visualise the count of paper submissions for journal articles of natural and social sciences |
| | [24] | – To animate networks over time |
| Noack algorithm [25] | [26] | − To visualise the land and water networks of port transportation |
| FA2 algorithm [27] | [28] | − To visualise the transaction patterns of Bitcoin networks |
| Hachul algorithm [29] | [30] | − To use $k$-dimensional trese (a data structure of space partitioning for arranging nodes in a $k$-dimensional space) to visualise networks. |

### 3.4 FORCE-DIRECTED ALGORITHMS IN BIOLOGICAL NETWORK VISUALISATION

Visualisation is an important way to capture the dependencies and interactions between different biological entities, and their sequential processes. The force-directed algorithm is one of the most popular approaches for the visualisation of biological networks. An example visualisation of a biological network is illustrated in Figure 7 (a) (which is generated by using the NGL molecular visualisation viewer [167]). Kerpedjiev et al. [168] developed a tool called *forna* to display the secondary structure of ribonucleic acid (RNA). In addition, Bang et al. [169] and Tuikkala et al. [170] proposed multilevel force-directed algorithms to visualise large protein networks and genetic interactions.

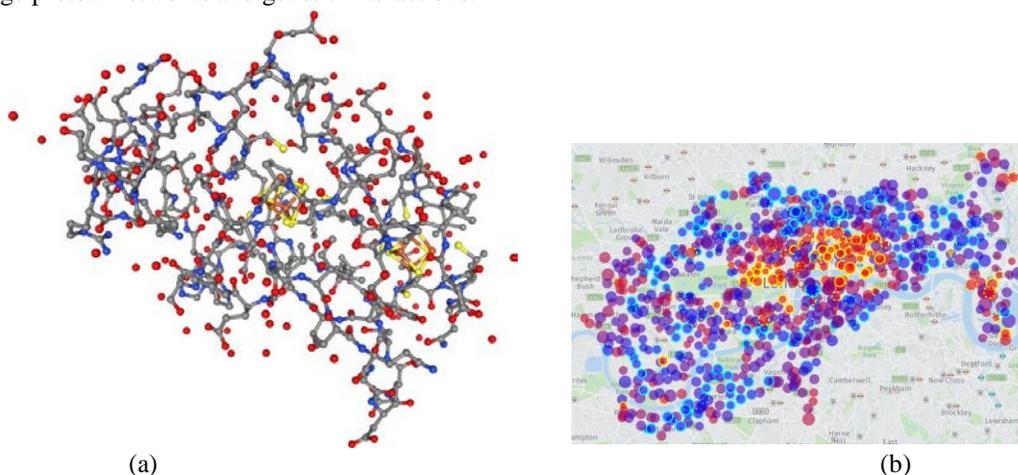

(a)                                                                      (b)

Figure 7 An example visualisation of (a) a biological network, (b) sensor localisation.

Biological networks have more special attributes than average directed and undirected networks. Because of this, various researchers have proposed special layouts for the visualisation of genetic sequencing or other biological networks. Clustered layouts are commonly used to visualise protein interactions [171]. Gamma-Clustering layouts were suggested for visualising large and complex biological networks [172]. Haplotype layouts [173] were also used to distinguish relationships among different sequences observed in biological networks [174]. There are also several studies that adopt force-directed algorithms to visualise the structure of molecules, biological pathways, protein networks, etc. Due to the page limitation, we summarise these studies in Table 3.

Table 3 Visualisation studies that adopted force-directed algorithms for biological networks.

| Force-directed algorithm adopted by the study | Study | Objective of the study |
|---|---|---|
| KK algorithm | [175] | − To visualise protein–protein interaction network |
| | [178] | − To visualise protein–protein interaction network |
| | [185] | − To visualise the structure of Alpha-helical transmembrane proteins |

| | | | |
|---|---|---|---|
| FR algorithm | [176] | – | To use Schlegel diagrams [177] to visualise the structure of molecules |
| | [179], [180], [181] | – | To visualise biological pathways |
| | [182] | – | To visualise the structure of genes |
| | | – | Minimise edge-edge crossings |
| | [183] | – | To visualise the structure of genes |
| | [184] | – | To visualise microarrays |
| | [186] | – | To visualise biological pathways |
| | [187] | – | To analyse the connectivity patterns of brain parcellation |
| | [172] | – | To visualise large biological networks |

## 3.5 FORCE-DIRECTED ALGORITHMS IN NODE PLACEMENT AND LOCALISATION FOR SENSOR NETWORKS

Sensor networks are useful for monitoring animals, earthquakes and tsunamis [188], emergency message forwarding during disasters [189], etc. An example visualisation of sensor localization is illustrated in Figure 5 (b) (which is generated by using the OOMap service [190]). Because the exact location of the networked sensors (nodes) is often unavailable, force-directed algorithms are used to determine node placement or to locate boundaries to improve the network's coverage [191]. The strength of force is subject to the distance between two nodes and each node behaves as a source of force. Therefore, if the distance between two nodes is shorter/larger than a threshold, a repulsive/attractive force will be exerted on each other. If the distance is equal to the threshold, no force will act upon the nodes. Extensions of the FR algorithm were proposed for node placement in [192, 193]. There are also extensions based on a modified FR algorithm that estimate the approximate location of each node based on signal information [194, 195]. In [196], Cheong and Si proposed a heuristic KK algorithm for boundary detection. The proposed algorithm was optimised for sending emergency messages via Mobile Ad Hoc network if cellular networks are corrupted. Nodes on the boundary are responsible for forwarding emergency messages to nearly emergency stations.

## 4 CONCLUSIONS

In this paper, we present the survey of force-directed algorithms for schematic drawings and placement. This class of algorithms has been studied and implemented in biological network visualisation, information visualisation, sensor localisation and VLSI design. This survey covers classical force-directed algorithms and hybrid force-directed algorithms, in which parallel, multilevel and multidimensional scaling techniques are used. We also discussed the merits and deficiencies of force-directed algorithms and visualisation applications. For example, how network topologies are drawn can significantly affect viewers' understanding of the network. We also discussed the influences caused by the layout and position-assignment of visualised network nodes on how a user perceives the relationships in the network. To this end, we review and categorise force-directed algorithms from research areas such as: (a) aesthetic drawings for general networks, (b) component placement and scheduling in high-level synthesis of very-large scale integration (VLSI) circuits design, (c) information visualisation, (d) biological network visualisation, and (e) node placement and localisation for sensor networks. Our hope is that this survey not only provides an overview of existing force-directed algorithms, but also introduces them as effective tools for solving visualisation problems in different application areas.

# APPENDIX

| Year | Aesthetic drawings for general networks | Biological network visualisation | Node placement and localisation for sensor networks | Information visualisation | Component placement in VLSI circuits design | Scheduling in VLSI circuits design |
|---|---|---|---|---|---|---|
| 1963 | 1 | 0 | 0 | 0 | 0 | 0 |
| 1964 | 0 | 0 | 0 | 0 | 0 | 0 |
| 1965 | 0 | 0 | 0 | 0 | 2 | 0 |
| 1966 | 0 | 0 | 0 | 0 | 0 | 0 |
| 1967 | 0 | 0 | 0 | 0 | 0 | 0 |
| 1968 | 0 | 0 | 0 | 0 | 0 | 0 |
| 1969 | 0 | 0 | 0 | 0 | 0 | 0 |
| 1970 | 0 | 0 | 0 | 0 | 0 | 0 |
| 1971 | 0 | 0 | 0 | 0 | 2 | 0 |
| 1972 | 0 | 0 | 0 | 0 | 0 | 0 |
| 1973 | 0 | 0 | 0 | 0 | 0 | 0 |
| 1974 | 0 | 0 | 0 | 0 | 2 | 0 |
| 1975 | 0 | 0 | 0 | 0 | 0 | 0 |
| 1976 | 0 | 0 | 0 | 0 | 0 | 0 |
| 1977 | 0 | 0 | 0 | 0 | 0 | 0 |
| 1978 | 0 | 0 | 0 | 0 | 0 | 0 |
| 1979 | 0 | 0 | 0 | 0 | 2 | 0 |
| 1980 | 0 | 0 | 0 | 0 | 0 | 0 |
| 1981 | 0 | 0 | 0 | 0 | 2 | 0 |
| 1982 | 0 | 0 | 0 | 0 | 4 | 0 |
| 1983 | 0 | 0 | 0 | 0 | 2 | 0 |
| 1984 | 0 | 0 | 0 | 0 | 0 | 0 |
| 1985 | 0 | 0 | 0 | 0 | 2 | 0 |
| 1986 | 0 | 0 | 0 | 0 | 2 | 0 |
| 1987 | 0 | 0 | 0 | 0 | 2 | 6 |
| 1988 | 0 | 0 | 0 | 0 | 2 | 0 |
| 1989 | 2 | 0 | 0 | 0 | 0 | 12 |
| 1990 | 0 | 0 | 0 | 0 | 0 | 0 |
| 1991 | 1 | 0 | 0 | 0 | 2 | 18 |
| 1992 | 0 | 0 | 0 | 0 | 0 | 6 |
| 1993 | 0 | 0 | 0 | 0 | 0 | 0 |
| 1994 | 2 | 0 | 0 | 0 | 0 | 0 |
| 1995 | 7 | 5 | 0 | 0 | 0 | 6 |
| 1996 | 3 | 0 | 0 | 0 | 0 | 6 |
| 1997 | 2 | 0 | 0 | 0 | 0 | 0 |
| 1998 | 1 | 0 | 0 | 0 | 2 | 0 |
| 1999 | 3 | 5 | 0 | 0 | 0 | 6 |
| 2000 | 4 | 0 | 0 | 0 | 0 | 12 |
| 2001 | 8 | 5 | 0 | 0 | 4 | 0 |
| 2002 | 3 | 0 | 0 | 0 | 4 | 6 |
| 2003 | 1 | 10 | 0 | 0 | 6 | 0 |
| 2004 | 5 | 10 | 4 | 0 | 4 | 0 |
| 2005 | 7 | 0 | 0 | 3 | 1 | 6 |
| 2006 | 4 | 5 | 0 | 3 | 6 | 6 |
| 2007 | 2 | 10 | 8 | 0 | 4 | 12 |
| 2008 | 4 | 10 | 0 | 15 | 4 | 0 |
| 2009 | 3 | 0 | 0 | 12 | 0 | 0 |
| 2010 | 5 | 15 | 4 | 0 | 2 | 0 |
| 2011 | 3 | 10 | 0 | 9 | 2 | 0 |
| 2012 | 6 | 5 | 4 | 9 | 4 | 6 |
| 2013 | 6 | 10 | 0 | 18 | 0 | 6 |
| 2014 | 5 | 5 | 0 | 12 | 0 | 0 |
| 2015 | 8 | 5 | 0 | 15 | 4 | 0 |
| 2016 | 6 | 5 | 4 | 24 | 2 | 0 |
| 2017 | 1 | 0 | 0 | 3 | 2 | 0 |

Figure 8 paper submission count of force-directed algorithms classified by application fields.